\def\year{2024}\relax
\documentclass[letterpaper]{article} 
\usepackage{aaai22}  
\usepackage{times}  
\usepackage{helvet}  
\usepackage{courier}  
\usepackage[hyphens]{url}  
\usepackage{graphicx} 
\usepackage{natbib}  
\usepackage{caption} 
\DeclareCaptionStyle{ruled}{labelfont=normalfont,labelsep=colon,strut=off} 
\usepackage{algorithm}
\usepackage{algorithmic}
\usepackage{booktabs} 

\usepackage{xcolor}
\definecolor{ForestGreen}{rgb}{0.13, 0.55, 0.13}

\usepackage{xcolor}

\usepackage{newfloat}
\usepackage{listings}
\floatstyle{ruled}
\newfloat{listing}{tb}{lst}{}
\floatname{listing}{Listing}

\title{Estimating Gender Completeness in Wikipedia}
\author {
    Hrishikesh Patel,
    Tianwa Chen,
    Ivano Bongiovanni,
    Gianluca Demartini
}
\affiliations {
    The University of Queensland, Brisbane, Australia\\
    \{hrishikeshm.patel, tianwa.chen, i.bongiovanni, g.demartini\}@uq.edu.au
}

\begin{document}

\maketitle

\begin{abstract}
Gender imbalance in Wikipedia content is a known challenge which the editor community is actively addressing.
The aim of this paper is to provide the Wikipedia community with instruments to estimate the magnitude of the problem for different entity types (also known as classes) in Wikipedia. To this end, we apply class completeness estimation methods based on the gender  attribute.
Our results show not only which gender for different sub-classes of Person is more prevalent in Wikipedia, but also an idea of how complete the coverage is for difference genders and sub-classes of Person.
\end{abstract}

\section{Introduction}
Successful crowdsourcing projects like Wikipedia and Wikidata naturally grow and evolve over time. This happens while having human editors focussing on certain parts of the project instead of others. While the ability for editors to decide what to contribute comes with the advantage of flexibility, it may result in biased content where, for example, one gender is better represented than others. An example of this is the number of male astronauts as compared to the number of female astronauts (76 female out of 630 astronauts in Wikipedia, as of the submission of this paper).

The editor community is actively addressing this \cite{langrock2022gender}.
Previous studies have shown how the editor population is also unbalanced from a gender point of view. \citet{antin2011gender} shows how the majority of editors, about 80\%, are male. Another important aspect to understand gender representation is that of measuring how many persons of a certain gender are represented by an article in Wikipedia. To this end, we look at instances of sub-classes of the class Person (e.g., astronauts). Such instances are represented by a single Wikipedia article (e.g, \url{https://en.wikipedia.org/wiki/Samantha_Cristoforetti}). The first step is that of counting how many instances (i.e., person) of a certain gender there are in such a class (e.g., astronaut) and make observations (e.g., 554 male astronauts and 76 female astronauts).

The more interesting step after counting entities is that of understanding how well represented each of the genders are in Wikipedia. To do this we would need to measure how many male/female astronauts there \textit{should be} in Wikipedia, assuming that the class is not yet completely represented in Wikipedia (e.g., because of a focus of the editor community on different parts of Wikipedia).

To close the gender gap in Wikimedia content it is first critical to be able to measure it.  To this end, in this paper we estimate the cardinality of sub-classes of Person based on gender. Similar to our approach, previous work by \citep{luggen_non-parametric_2019} has used statistical estimators for class cardinality using Wikidata edit history in a capture/recapture setup. In a similar fashion, in our work we apply such estimators but using the Wikipedia edit history and also taking a gender-based approach to it.
This allows us to not only estimate the cardinality of a class (e.g., female astronauts), but also, by comparing the estimated size with the  number of male/female instances currently present in Wikipedia, to measure the \textit{completeness} of each class for different genders (e.g., male astronauts are 95\% complete while female astronauts are 94\% complete), which is our research question.

\section{Related Work}

\subsection{Gender in Wikipedia}
Previous research has already looked at the gender gap \cite{farzan2016bring}, which is a particularly important issue across Wikimedia projects \cite{redi2020taxonomy}. For example, \citet{abian2022analysis} proposed methods to identify content gaps in crowdsourced knowledge graphs and found that Wikidata editors usually tend not to work on under-represented entities. While many attempts to address this issue exist, the key difference in the approach we propose to take is that we do not attempt to address the issue, but rather, to apply effective methods and develop accurate instruments for the different Wikipedia user groups to be empowered in making data-driven editorial decisions and able to address the issue by themselves.

\subsection{Class Cardinality Estimation}
Previous research \cite{luggen_non-parametric_2019} has looked at how to use statistical estimators to estimate class cardinality in Wikidata. They used the knowledge graph edit history as evidence for the estimators. In this paper we extend this approach by looking at attribute-specific cardinality estimations (e.g., How many female astronauts should be there? Do we have them all?) and beyond the Wikidata project.

In the area of crowdsourced databases, the problem of answering queries under the open world assumption has been studied in the past. Researchers have encountered the problem that popular entities are reported by crowd members more frequently than ``tail'' (i.e., unpopular) entities, thus making it difficult to complete the answer set (and, in our case, to estimate the class cardinality). The approach followed by \citet{trushkowsky2014crowdsourcing} has looked at using statistical estimators to understand how far from the complete set the incrementally constructed query answer set currently is. We apply these methods to understand how complete Wikipedia is.

\section{Data Collection}\label{sec:data}
The aim is to generate a dataset that enables us to estimate gender-based class cardinality first and  completeness levels next. To this end, we first collected the list of sub-classes (e.g., Artist, Astronaut, Monarch, etc.) of the class `Person' from DBpedia\footnote{\url{https://www.dbpedia.org/about/}} as well as the list of all entities assigned to each of these sub-classes. For each of the collected entities, we then retrieved its Wikipedia article edit history from the MediaWiki API for the period from Jan 2019 to Dec 2023 to be used as ``capturing'' events for the statistical estimators.
%
In total, our dataset contains  121,535 entities over 34 classes and a total of 4,896,299 edits. The created dataset can be found online at [Redacted for blind review].
The size of the dataset is 1.2GB and was processed on a 32GB RAM server.

\section{Name-based Gender Estimation}
Based on previous work by \citet{van2023open}, we first estimate the gender of persons described by the Wikipedia articles.  We used their 'nomquamgender' package\footnote{\url{https://github.com/ianvanbuskirk/nomquamgender/}}, available under the MIT license.
Using the dataset described above, we perform name-based gender classification for each of the collected entities of type `Person'. 
This enables us to make initial observations about the gender distribution of persons having an article in Wikipedia for different sub-classes of Person. 

\begin{figure}[ht]
    \centering
    \includegraphics[width=0.35\textwidth]{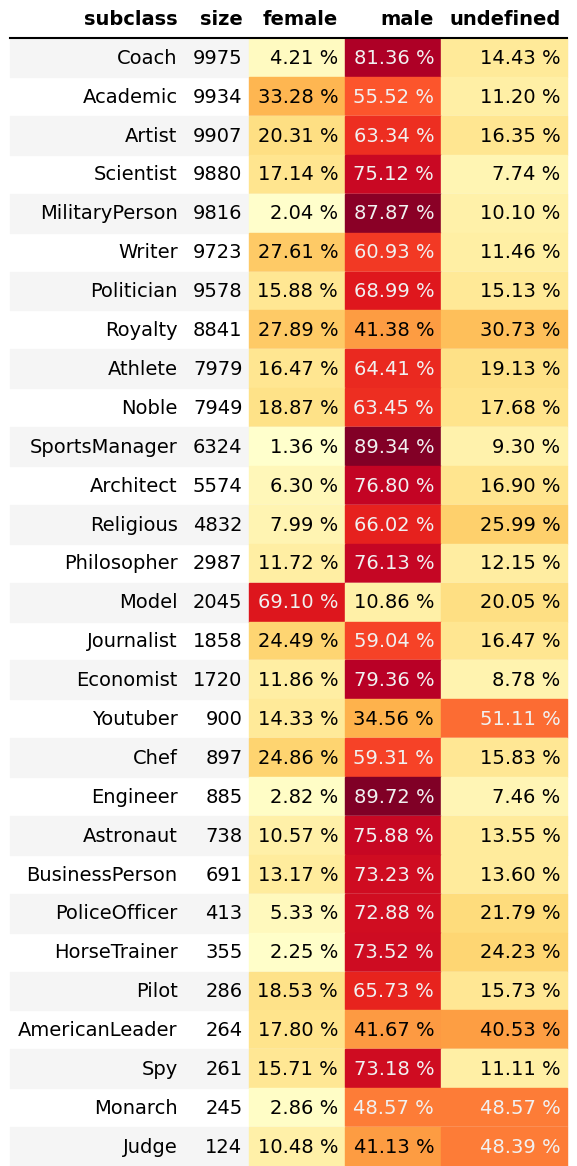}
    \caption{Name-based gender classification of Wikipedia person names per sub-class of Person.}
    \label{fig:genderclassification}
\end{figure}

Figure \ref{fig:genderclassification} shows the results of this classification task\footnote{The approach would not classify a name if its confidence is low and thus we keep a column for undefined gender in Fig. \ref{fig:genderclassification}}.
We can observe how, unsurprisingly, certain sub-classes of Person are less gender balanced than others (e.g., `Economist' being mostly male and `Model' being mostly female).

\section{Gender-based Class Cardinality Estimation}\label{sec:methods}
\begin{figure*}
    \includegraphics[width=\textwidth]{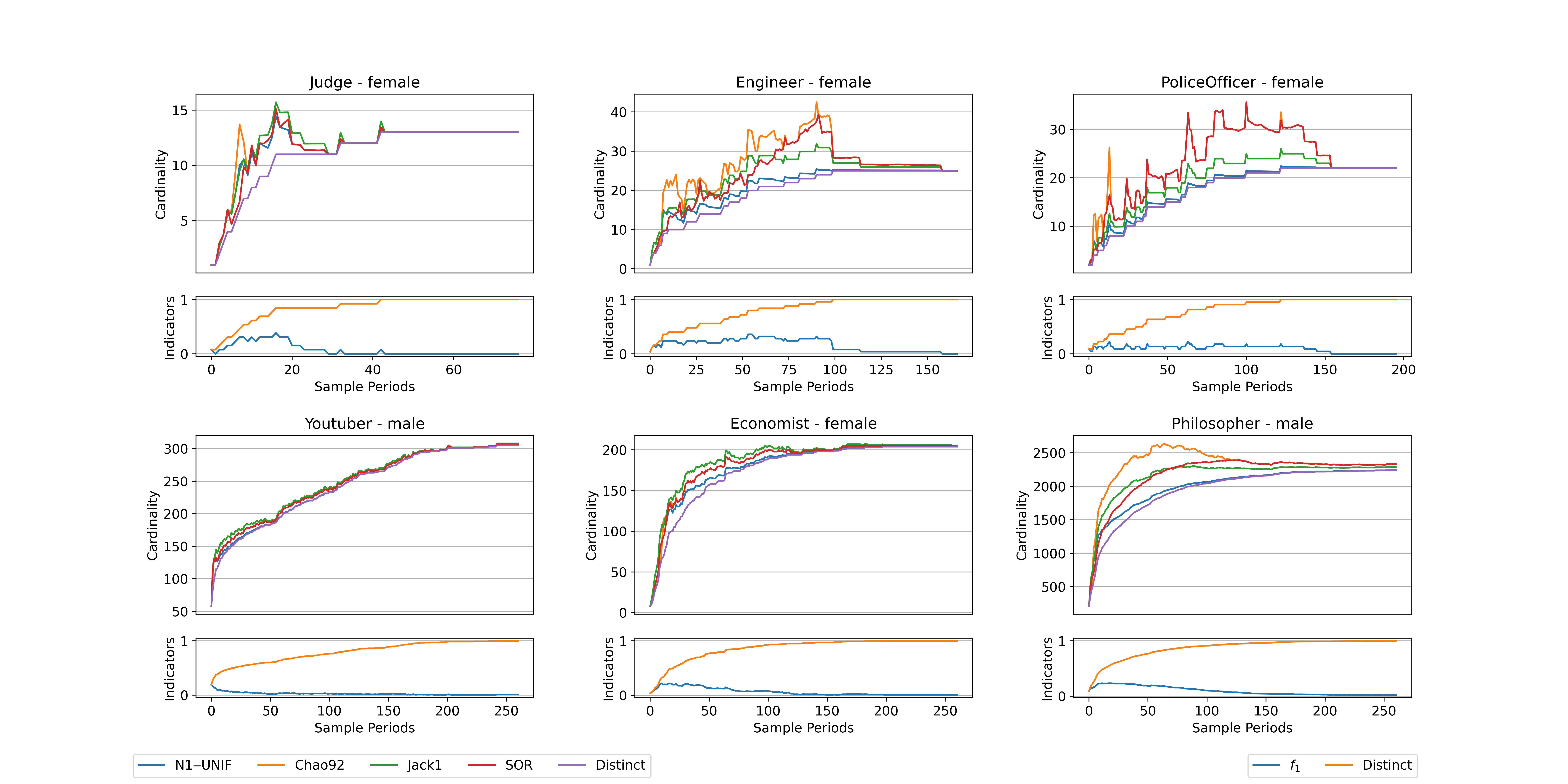}
    \caption{Class cardinality estimation for some example classes (classes estimated to be most incomplete in the top row, and classes estimated to be most complete in the bottom row) with different statistical estimators.}
    \label{fig:estimations}
\end{figure*}

Once we have classified the gender of the people with a Wikipedia article, we can now proceed with the estimation of the cardinality of each sub-class of Person (i.e., how many Astronauts should we have in Wikipedia) using the edit history of their articles.
In short, these methods make use of samples of entities from a population (e.g., Astronauts) to estimate the size of the population. The capture/recapture methods we use, originally designed in computational ecology (e.g., capturing and tagging lions in the Savanna to estimate the size of the entire population of lions), require a concept of `sampling' (i.e., capturing samples of the population over time). In our setting, we make use of the edit of a Wikipedia article as the sampling event. More popular entities will receive more edits (and thus will be sampled more often) than less popular ones. With this data, the more samples we have the more accurate the estimators will be, eventually converging to the true value of the population size.
The other good aspect of these estimators is that they provide a confidence score that tell us how accurate the estimation is.
More details on the statistical estimation methods we use can be found in \cite{luggen_non-parametric_2019}.
%
Some example population size estimation results are presented in Figure \ref{fig:estimations}.
These methods allow us to compute how many entities of a certain type there \textit{should} be in Wikipedia and, with that, compute an estimated completeness level by comparing the estimanted cardinality with the current number of entities of that class in Wikipedia.

\subsection{Window Size Parameter}
\begin{figure}[t]
    \centering
    \includegraphics[width=0.55\textwidth]{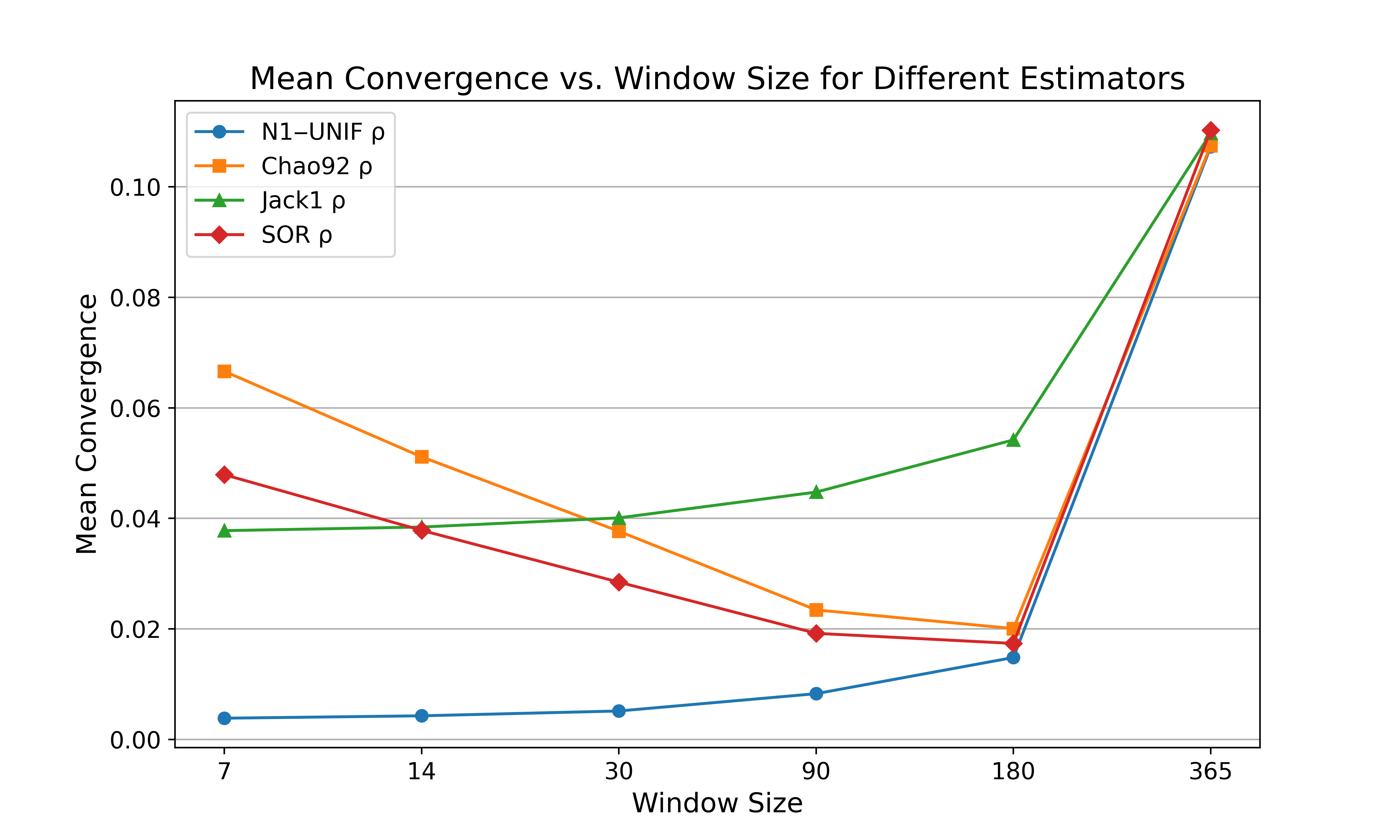}
    \caption{The impact of window size over the edit history for capture/recapture-based estimation methods.}
    \label{fig:windowsize}
\end{figure}

One open parameter we look at is that of the size of the window (i.e., interval of time) over which we observe capture/recapture events (i.e., Wikipedia article edits).
We consider an entity being `captured' if it has been edited in the period of time, so in a one year window almost all entities are expected to appear.
Figure \ref{fig:windowsize} shows convergence score (the lower the better) for different window  sizes (one and two weeks; one, three, and six months; and one year).
Results indicate that 7 days is best for both Jack1 (J1) and N1\_UNIF (N1) estimators. In the following, we use this window size for calculating our estimates.

\subsection{Cardinality and Completeness Results}
\begin{figure*}
    \centering
    \includegraphics[width=0.7\textwidth]{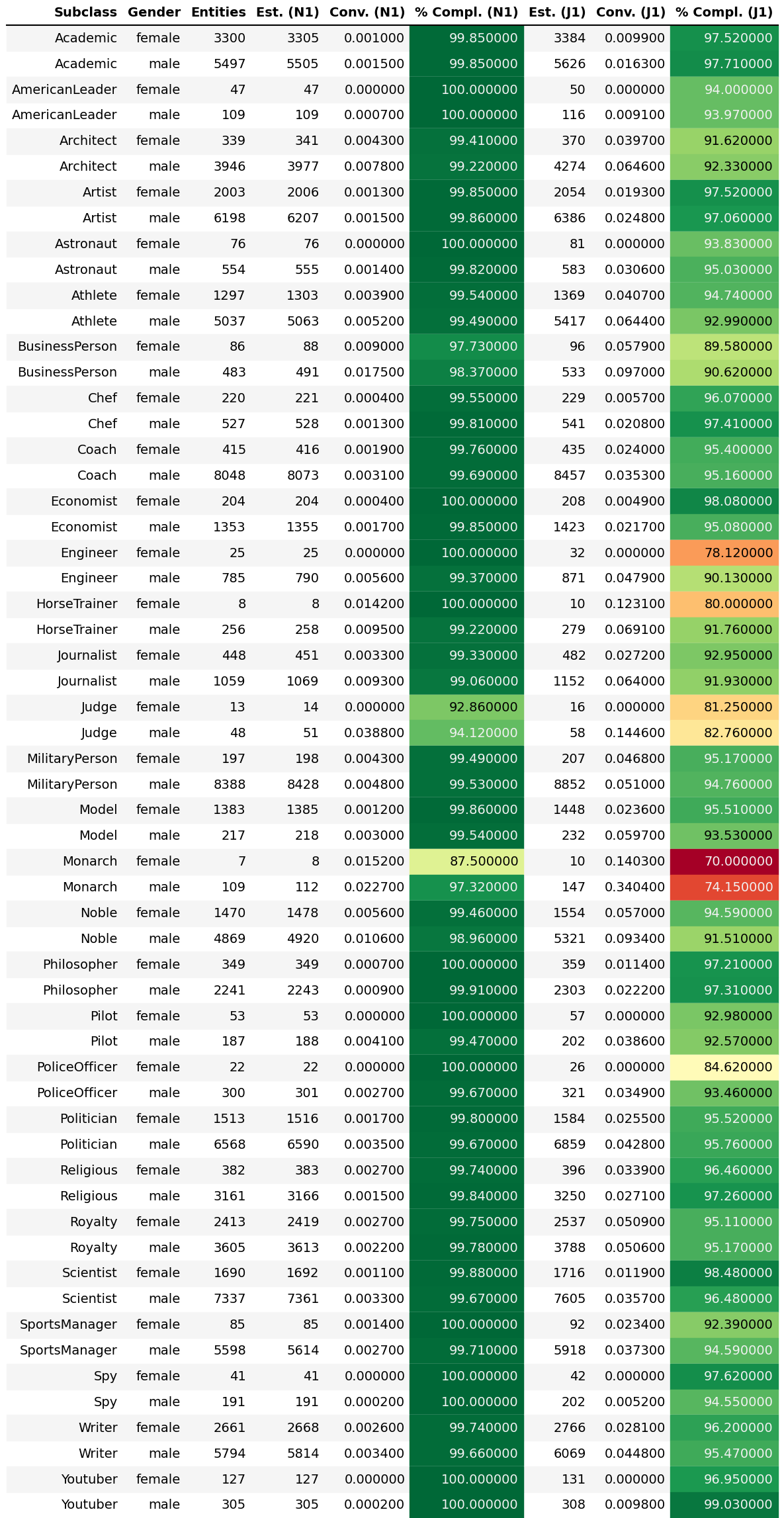}
    \caption{Gender-based entity count in Wikipedia, cardinality estimation (Est.), convergence score (Conve.), and estimated class completeness (Compl.) for estimators N1\_UNIF (N1) and Jack1 (J1).}
    \label{fig:results_table}
\end{figure*}

As J1 is the best estimator according to \citet{luggen_non-parametric_2019} and N1 was also good according to our convergence analysis (Fig. \ref{fig:windowsize}),
we select these as the estimators to use. 
Fig. \ref{fig:results_table} shows current size, estimated class cardinality, and completeness level for sub-classes of Person split over two gender values.

\section{Discussion and Conclusions}\label{sec:conc}
Figure \ref{fig:results_table} shows some informative data like, for example, the class of female engineers only being 78\% complete as compared to 90\% for male engineers (based on J1).
These are important observations as an under-representation of notable female engineers may reinforce stereotypes that successful engineers are man and may discourage certain young students targeting STEM careers.

Overall, we observe high completeness levels. This is consistent with the generally perceived high quality of Wikipedia content. Our estimates show that  16 sub-classes have higher completion rate for male and 13 have a higher completion rate for female.
Possible limitations include the use of DBpedia sub-classes of Person which may be imperfect, as well as the impossibility to assess estimation accuracy due to the lack of a ground truth.

Such methods and results may be useful to the Wikipedia editor community to inform editorial decision-making processes (e.g., given this data, the Wikipedia editor community may decide to stop adding new male engineers for a period of time to focus on only adding new female engineers).

\newpage
\bibliography{references,wikigender}

\end{document}